\documentstyle[12pt,twoside]{article}
 
\def\a{\alpha}
\def\b{\beta}
\def\c{\chi}
\def\d{\delta}
\def\e{\epsilon}                
\def\f{\phi}                    
\def\g{\gamma}
\def\h{\eta}

\def\k{\kappa}
\def\l{\lambda}
\def\m{\mu}
\def\n{\nu}

\def\p{\pi}                     
\def\r{\rho}                    
\def\s{\sigma}                  
\def\t{\tau}

\def\F{\Phi}
\def\G{\Gamma}

\def\P{\Pi}

\def\cd{{\cal D}}

\def\cl{{\cal L}}

\def\bo{{\raise.05ex\hbox{\large$\Box$}\:}}             
\def\cbo{{\,\raise-.15ex\Sc [\,}}                       
\def\pa{\partial}                                       
\def\su{\sum}                                           
\def\TH{{\raise.2ex\hbox{$\displaystyle \bigodot$}\mskip-4.7mu \llap H \;}}
\def\face{\hbox{\normalsize$\;\;\:{\raise.9ex\hbox{\oo n}\mskip-13mu \llap
        {${\buildrel{\hbox{\frtnrm ..}}\over\smile}$}}\:$}}     
\def\Face{{\raise.2ex\hbox{$\displaystyle \bigodot$}\mskip-2.2mu \llap {$\ddot
        \smile$}}}                                      
\def\Lhat{{\bf\rlap{\kern-.09em$\hat{\phantom L}$}L}}
\def\Lcheck{{\bf\rlap{\kern-.09em$\check{\phantom L}$}L}}
 
\def\sp#1{{}^{#1}}                              
\def\sb#1{{}_{#1}}                              

\def\leftrightarrowfill{$\mathsurround=0pt \mathord\leftarrow \mkern-6mu
        \cleaders\hbox{$\mkern-2mu \mathord- \mkern-2mu$}\hfill
        \mkern-6mu \mathord\rightarrow$}
\def\dvec#1{\vbox{\ialign{##\crcr
        \leftrightarrowfill\crcr\noalign{\kern-1pt\nointerlineskip}
        $\hfil\displaystyle{#1}\hfil$\crcr}}}           
\def\ddt#1{{\buildrel {\hbox{\LARGE .\kern-2pt.}} \over {#1}}}
 
\def\frac#1#2{{\textstyle{#1\over\vphantom2\smash{\raise.20ex
        \hbox{$\scriptstyle{#2}$}}}}}                   
\def\sfrac#1#2{{\vphantom1\smash{\lower.5ex\hbox{\small$#1$}}\over
        \vphantom1\smash{\raise.4ex\hbox{\small$#2$}}}} 
\def\bfrac#1#2{{\vphantom1\smash{\lower.5ex\hbox{$#1$}}\over
        \vphantom1\smash{\raise.3ex\hbox{$#2$}}}}       
\def\afrac#1#2{{\vphantom1\smash{\lower.5ex\hbox{$#1$}}\over#2}}    
 
\def\boxes#1{
        \newcount\num
        \num=1
        \newdimen\downsy
        \downsy=-1.64ex
        \mskip-7.8mu
        \bo
        \loop
        \ifnum\num<#1
        \llap{\raise\num\downsy\hbox{$\bo$}}
        \advance\num by1
        \repeat}
\def\boxup#1#2{\newcount\numup
        \numup=#1
        \advance\numup by-1
        \newdimen\upsy
        \upsy=.82ex
        \mskip7.8mu
        \raise\numup\upsy\hbox{$#2$}}
 
\newskip\humongous \humongous=0pt plus 1000pt minus 1000pt
\def\caja{\mathsurround=0pt}

\newif\ifdtup
\def\panorama{\global\dtuptrue \openup2\jot \caja
        \everycr{\noalign{\ifdtup \global\dtupfalse
        \vskip-\lineskiplimit \vskip\normallineskiplimit
        \else \penalty\interdisplaylinepenalty \fi}}}
\def\li#1{\panorama \tabskip=\humongous                         
        \halign to\displaywidth{\hfil$\displaystyle{##}$
        \tabskip=0pt&$\displaystyle{{}##}$\hfil
        \tabskip=\humongous&\llap{$##$}\tabskip=0pt
        \crcr#1\crcr}}

\def\NP{Nucl. Phys. B\,}

\def\PRD{Phys. Rev. D\,}
\def\CQG{Class. Quant. Grav.}

\def\ref#1{$\sp{#1]}$}

\parskip=\medskipamount                 
\def\baselinestretch{1.2}       
\thicklines                         
\def\title#1#2#3#4{
\begin{document}
        {\hbox to\hsize{#4 \hfill  #3}}\par
        \begin{center}\vskip.5in minus.1in {\Large\bf #1}\\[.5in minus.2in]{#2}
        \vskip1.4in minus1.2in {\bf ABSTRACT}\\[.1in]\end{center}
        \begin{quotation}\par}
\def\author#1#2{#1\\[.1in]{\it #2}\\[.1in]}

\def\AMIC{Aleksandar Mikovic\'c
\\[.1in]{\it Blackett Laboratory, Imperial College, Prince Consort Road, London
SW7 2BZ, UK}\\[.1in]}

\def\AMICIF{Aleksandar Mikovi\'c\,
\footnote{Work supported by MNTRS and Royal Society}
\\[.1in] {\it Blackett Laboratory, Imperial College, Prince Consort
Road, London SW7 2BZ, UK}\\[.1in]
and \\[.1 in]
{\it Institute of Physics, P.O. Box 57, 11001 Belgrade, Yugoslavia}
\footnote{Permanent address}\\ {\it E-mail:\, mikovic@castor.phy.bg.ac.yu}}

\def\AMSISSA{Aleksandar Mikovi\'c\,
\footnote{E-mail address: mikovic@castor.phy.bg.ac.yu}
\\[.1in] {\it SISSA-International School for Advanced Studies\\
Via Beirut 2-4, Trieste 34100, Italy}\\[.1in]
and \\[.1 in]
{\it Institute of Physics, P.O. Box 57, 11001 Belgrade, Yugoslavia}
\footnote{Permanent address}}

\def\AM{Aleksandar Mikovi\'c 
\footnote{E-mail address: mikovic@castor.phy.bg.ac.yu}
\\[.1in] {\it Institute of Physics, P.O.Box 57, Belgrade 11001, Yugoslavia}
\\[.1in]}

\def\AMsazda{Aleksandar Mikovi\'c 
\footnote{E-mail address: mikovic@castor.phy.bg.ac.yu}
and Branislav Sazdovi\'c \footnote{E-mail: sazdovic@castor.phy.bg.ac.yu}
\footnote{Work supported by MNTRS}
\\[.1in] {\it Institute of Physics, P.O.Box 57, Belgrade 11001, Yugoslavia}
\\[.1in]}

\def\AMVR{Aleksandar Mikovi\'c\,
\footnote{E-mail address: mikovic@castor.phy.bg.ac.yu}
\\[.1in] 
{\it Institute of Physics, P.O. Box 57, 11001 Belgrade, Yugoslavia}
\\[.2in]
Voja Radovanovi\'c \\[.1 in]
{\it Faculty of Physics, P.O. Box 550, 11001 Belgrade, Yugoslavia}}

\def\AMCVR{Aleksandar Mikovi\'c
\footnote{Permanent address: Institute of Physics, P.O. Box 57, 11001 
Belgrade, Yugoslavia}\footnote{E-mail: mikovic@fy.chalmers.se, 
mikovic@castor.phy.bg.ac.yu}
\\
{\it Institute of Theoretical Physics, Chalmers University of Technology,
S-412 96 Goteborg, Sweden}\\[.1in]
and
\\[.1in]
Voja Radovanovi\'c
\footnote{E-mail: rvoja@rudjer.ff.bg.ac.yu} \\
{\it Faculty of Physics, P.O. Box 550, 11001 Belgrade, Yugoslavia}}

\def\AMVVR{Aleksandar Mikovi\'c
\footnote{On leave from Institute of Physics, P.O. Box 57, 11001 
Belgrade, Yugoslavia}
\footnote{Supported by Comissi\'on Interministerial de Ciencia y Tecnologia}
\footnote{E-mail: mikovic@lie1.ific.uv.es}
\\
{\it Departamento de Fisica Te\'orica and IFIC, Centro Mixto Universidad
de Valencia-CSIC, Facultad de Fisica, Burjassot-46100, Valencia, Spain}
\\[.1in]
Voja Radovanovi\'c
\footnote{E-mail: rvoja@rudjer.ff.bg.ac.yu} \\
{\it Faculty of Physics, P.O. Box 368, 11001 Belgrade, Yugoslavia}}

\def\AMV{Aleksandar Mikovi\'c
\footnote{On leave from Institute of Physics, P.O. Box 57, 11001 
Belgrade, Yugoslavia}
\footnote{Supported by Comissi\'on Interministerial de Ciencia y Tecnologia}
\footnote{E-mail: mikovic@lie1.ific.uv.es}
\\
{\it Departamento de Fisica Te\'orica and IFIC, Centro Mixto Universidad
de Valencia-CSIC, Facultad de Fisica, Burjassot-46100, Valencia, Spain}}

\def\endtitle{\par\end{quotation}\vskip3.5in minus2.3in\newpage}
 
 
\def\endabstract{\par\end{quotation}
        \renewcommand{\baselinestretch}{1.2}\small\normalsize}
 
 
\def\xpar{\par}                                         

\def\letterhead{
        \centerline{\large\sf INSTITUTE OF PHYSICS}
        \centerline{\sf P.O.Box 57, 11001 Belgrade, Yugoslavia}
        \rightline{\scriptsize\sf Dr Aleksandar Mikovi\'c}
        \vskip-.07in
        \rightline{\scriptsize\sf Tel: 11 615 575}
        \vskip-.07in
        \rightline{\scriptsize\sf E-mail: MIKOVIC@CASTOR.PHY.BG.AC.YU}}

\def\sig#1{{\leftskip=3.75in\parindent=0in\goodbreak\bigskip{Sincerely yours,}
\nobreak\vskip .7in{#1}\par}}

\def\ssig#1{{\leftskip=3.75in\parindent=0in\goodbreak\bigskip{}
\nobreak\vskip .7in{#1}\par}}

 
\def\ree#1#2#3{
        \hfuzz=35pt\hsize=5.5in\textwidth=5.5in
        \begin{document}
        \ttraggedright
        \par
        \noindent Referee report on Manuscript \##1\\
        Title: #2\\
        Authors: #3}
 
 
\def\start#1{\pagestyle{myheadings}\begin{document}\thispagestyle{myheadings}
        \setcounter{page}{#1}}
 
 
\catcode`@=11
 
\def\ps@myheadings{\def\@oddhead{\hbox{}\footnotesize\bf\rightmark \hfil
        \thepage}\def\@oddfoot{}\def\@evenhead{\footnotesize\bf
        \thepage\hfil\leftmark\hbox{}}\def\@evenfoot{}
        \def\sectionmark##1{}\def\subsectionmark##1{}
        \topmargin=-.35in\headheight=.17in\headsep=.35in}
\def\ps@acidheadings{\def\@oddhead{\hbox{}\rightmark\hbox{}}
        \def\@oddfoot{\rm\hfil\thepage\hfil}
        \def\@evenhead{\hbox{}\leftmark\hbox{}}\let\@evenfoot\@oddfoot
        \def\sectionmark##1{}\def\subsectionmark##1{}
        \topmargin=-.35in\headheight=.17in\headsep=.35in}
 
\catcode`@=12
 
\def\sect#1{\bigskip\medskip\goodbreak\noindent{\large\bf{#1}}\par\nobreak
        \medskip\markright{#1}}
\def\chsc#1#2{\phantom m\vskip.5in\noindent{\LARGE\bf{#1}}\par\vskip.75in
        \noindent{\large\bf{#2}}\par\medskip\markboth{#1}{#2}}
\def\Chsc#1#2#3#4{\phantom m\vskip.5in\noindent\halign{\LARGE\bf##&
        \LARGE\bf##\hfil\cr{#1}&{#2}\cr\noalign{\vskip8pt}&{#3}\cr}\par\vskip
        .75in\noindent{\large\bf{#4}}\par\medskip\markboth{{#1}{#2}{#3}}{#4}}
\def\chap#1{\phantom m\vskip.5in\noindent{\LARGE\bf{#1}}\par\vskip.75in
        \markboth{#1}{#1}}
\def\refs{\bigskip\medskip\goodbreak\noindent{\large\bf{REFERENCES}}\par
        \nobreak\bigskip\markboth{REFERENCES}{REFERENCES}
        \frenchspacing \parskip=0pt \renewcommand{\baselinestretch}{1}\small}
\def\unrefs{\normalsize \nonfrenchspacing \parskip=medskipamount}
\def\Item{\par\hang\textindent}
\def\Itemitem{\par\indent \hangindent2\parindent \textindent}
\def\makelabel#1{\hfil #1}
\def\topic{\par\noindent \hangafter1 \hangindent20pt}
\def\Topic{\par\noindent \hangafter1 \hangindent60pt}

\title{One-loop Effective Action for a Generic 2D Dilaton Gravity Theory}
{\AMVVR}{FTUV/97-17, IFIC/97-17}{April 1997}

We study the one-loop effective action for a generic two-dimensional dilaton 
gravity theory conformally coupled to $N$ matter fields. We obtain an
explicit expression for the effective action in the weak-coupling limit
under a suitable restriction of the dilaton potential asymptotics.
Our result applies to the CGHS model as well as to the spherically
symmetric general relativity. The effective action is obtained by
using the background-field method, and we take into account the loop 
contributions from all the fields in the classical action and from the ghosts.  
In the large-$N$ limit, and after an appropriate field redefinition, 
the one-loop correction takes the form of the Polyakov-Liouville action.
    
\endtitle

\sect{1. Introduction}

Two-dimensional (2d) dilaton gravity theories are useful models for gaining
understanding about the quantum properties of black holes \cite{s}.
A usual way of exploring the quantum effects is by studying the effective 
action, which can be obtained by using the standard covariant
perturbation techniques. Covariant perturbative
quantization of 2d dilaton gravity theories has been studied
by many authors \cite{gkt,kn,os1,os2,ea,kkn,kst,bss}, 
but surprisingly, there has not been any systematic
study of the effective action. In the context of 2d sigma models,
which are related to 2d dilaton gravities, a one-loop contribution due to
the scalar fields has been evaluated in \cite{ht}. In 
the case of the CGHS model \cite{cghs}, an incomplete one-loop effective action
has been found
from a combination of the path-integral measure and the $\b$-function
considerations \cite{dea,gs}. Subsequently, further one-loop terms  
were found from the symmetry considerations \cite{rst,bpp}. 

In order to give a systematic derivation of the one-loop CGHS effective action,
and in order to do the same for a more realistic model of the spherically 
symmetric general relativity,
we will study the one-loop effective action of a generic dilaton gravity theory
coupled to conformal matter. The classical action of such a theory is given by
$$ S =  \int d^2 x \sqrt  {\tilde g }\left[ e^{-2\F}\left( \tilde R
 + \a ( \tilde \nabla \F)^2 + U(\F) \right) - {1\over 2}\su_{i}
( \tilde \nabla f_i )^2 \right]\quad ,
\eqno(1.1)$$
where $\tilde g$, $\tilde R$ and $\tilde \nabla$ stand for the 
determinant,
the scalar curvature and the covariant derivative associated to the 
physical 2d metric ${\tilde g}_{\m\n}$, $\a$ is a numerical constant
and $i=1,...,N$. 
$\F$ is the dilaton scalar field, while $f_i$ are the
matter scalar fields.
When $\a =4$ and $U(\F) =$const.$=4\l\sp 2$ the action (1.1) is the CGHS model,
while $\a =2$ and $U(\F) = e\sp{2\F}$ gives the spherically
symmetric general relativity coupled to null-dust (in units $G=1$, where
$G$ is the Newton constant)
\cite{cmn}, to which we refere as the SSND model. 
If the dilaton is redefined as $\f = e\sp{-2\F}$, and after an
appropriate rescaling of the metric 
(${\tilde g}\sb{\m\n} = e\sp{\a\F/2}g\sb{\m\n}$), 
the action (1.1) simplifies 
$$ S =  \int d^2 x \sqrt  { g }\left[ \f R
 + V(\f) - {1\over 2}\sum_{i}( \nabla f_i )^2 \right]\quad .
\eqno(1.2)$$
In this form the CGHS model is given by $V(\f) =4\l\sp 2$, while 
the SSND model is given by $V(\f) = 1/\sqrt{\f}$.
The form (1.2) is more convenient for the calculation of the effective action,
and the only restriction on the dilaton potential will be that for large $\f$
the potential $V$ 
will behave as $\f\sp{-k}$, where $k\ge 0$. This ensures that in the 
weak-coupling limit $\f\to \infty$ (which for the spherically symmetric
general relativity means large radius $r$, since in that case $\f =r\sp 2$
\cite{cmn}) we can ignore the contributions to the
effective action proportional to $d\sp n V/d\f\sp n$ or to $V\sp{n+1}$ 
for $n\ge 1$.
In this limit we will calculate the complete one-loop effective action,
which will include the loop contributions
from the metric, the dilaton and the ghosts. 
 
In section 2 we set up the background field method we will use for the
calculation of the one-loop effective action. In section 3 we expand the
results of section 2 to the first order in a perturbation of the background
metric, since this simplifies the calculations. In section 4 we calculate the
contribution to the effective action which is independent of $\f$-derivatives.
In section 5 we calculate the contribution for a flat background metric.
In section 6 we combine the results of sectons 4 and 5 and give our final
result. In section 7 we present our conclusions. 

\sect{2. The background field method}

In order to calculate the one-loop effective action
we will use the background field method in a formulation given by Abbot 
\cite{abbot}. The effective action $\G[\f_0]$ can be expressed as
$$
e^{i\G[\f _0]}=\int \cd\f e^{i(S(\f ) +\int d^2x J(\f-\f _0))}\quad,\eqno(2.1)
$$ 
where $\f_0$ denotes a set of the classical background fields.
$S$ is a sum of the classical action, the gauge-fixing action and
the ghost action, so that the integration variable $\f$ also includes
the ghost fields. However, for the ghosts $J=0$ and $\f_0 =0$.
The source $J$ is a function of $\f_0$, which is determined from
$$ J = -{\d \G\over\d\f_0} = -{\d S\over\d\f_0} + O(\hbar) \quad.$$
One then splits $\f$ as
$\f=\f _0+\tilde\f$ where $\tilde\f$ is a new integration variable,
called the quantum field. The path-integral (2.1) is then evaluated
perturbatively, by Taylor expanding $S(\f _0+{\tilde\f})$. For the one-loop
approximation, one expands $S$ to the second order in $\tilde\f$, 
while $J$ is expanded to the zeroth order in $\hbar$, so that
$$e^{i\G [\f_0]}\approx e^{iS(\f_0)} \int \cd {\tilde \f}
e^{{i\over 2} S^{\prime \prime}(\f _0){\tilde \f}^2}\ , \eqno(2.2)
$$
where $''$ denotes the second functional derivative. Approximation (2.2) then 
yields the one-loop effective action formula 
$$\G_1 [\f_0] =S(\f _0)-{1 \over 2i} {\rm Tr} 
\left(\log S^{\prime \prime }(\f _0)\right)\quad .
\eqno(2.3)$$

In our case, we split the fields $\{ g_{\m\n},\ \f,\ f\}$ into the classical
background fields $\{g_{\m\n},\  \f,\ f_0\}$ and the quantum fields
$\{h_{\m\n},\ \hat \f,\  f\}$ as
$$ g_{\m\n}\rightarrow g_{\m\n}+h_{\m\n} \quad,
\quad \f \rightarrow  \f+\hat\f \quad,
\quad f\rightarrow f_0+f \quad,\eqno(2.4)$$
so that quadratic in quantum fields part of the action (1.2) is given by
$$\li{ S^{(2)}=\int d^2x\sqrt g  &{\Big[}{1\over 4}\f h^{\m\n}\Box
h_{\m\n}+{1\over 4}D_{\r}\f hD^{\r}h-{3\over 4}
D_{\b}\f h^{\r\m}D^{\b}h^{\r\m}\cr
 &-D_{\n}\f D_{\r}hh^{\r\n}
-{1\over 2}\f h^{\n\b}D_{\b}D_{\m}h^{\m}_{\n}+D_{\m}\f h^{\r\m}D^{\n}h_{\r\n}+
{1\over 2}D_{\r}\f h^{\m}_{\b}D_{\m}h^{\r\b}+\cr 
& +{1\over 2}\f h D_{\m}D_{\n}h^{\m\n}
+\f h^{\m\l}h^{\n}_{\l}R_{\n\m} +{1\over 8}\f Rh^2-{1\over 4}\f h\Box
h-{1\over 4}\f Rh_{\m\n}h^{\m\n}\cr 
 &-{1\over 2}
   \f hh^{\m\n}R_{\m\n}+{1\over 2}h\hat \f R
   +  \hat \f D_{\m}D_{\n}h^{\m\n}-\hat \f\Box h-\hat \f h^{\m\n}R_{\m\n}\cr
&+ ({1\over 8}h^2-{1\over 4}h^{\m\n}h_{\m\n})V(\f ) +{1\over 2}
V^{\prime \prime }(\f )
\hat \f^2 + {1\over 2} V^{\prime}(\f )h\hat \f \cr
&-
{1\over 2}g^{\m\n}\pa _{\m}f\pa_{\n}f
-{1\over 2} \pa _{\m}f_0\pa _{\n}f_0(-{1\over 4}g^{\m\n}h_{\a\b}h^{\a\b}\cr
&-{1\over 2}
hh^{\m\n}+
 h^{\m\l}h_{\l}^{\n}+{1\over 8}g^{\m\n}h^2)-\pa _{\m}f\pa
_{\n}f_0({1\over 2}hg^{\m\n}-h^{\m\n} ){\Big]}\quad.& (2.5)\cr }
$$ 
The indices in (2.5) are lowered and raised by  the classical metric
$g_{\m\n}$, while $h=g^{\m\n}h_{\m\n}$ and $D_{\m} $  is the covariant
derivative associated to $g_{\m\n}$.
$S^{(2)}$ has two type of gauge symmetries; classical and quantum. 
This ensures that after fixing of the
quantum guage symmetry one still obtains a gauge-invariant effective action.
We choose the  gauge-fixing condition as 
$$\c _{\m}=D_{\l}h^{\l}_{\m}-{1\over 2}D_{\m}h -{1\over \f }D_{\m}\hat \f =0
\quad,\eqno(2.6)$$
so that the gauge-fixing term in the action takes the form
$$S_{GF}=-{1\over 2}\int dx\sqrt g\f \c _{\m}\c ^{\m}\quad.\eqno(2.7)$$
(2.7) is chosen such that the total action has a minimal structure, i.e. 
the second spacetime derivatives acting on the quantum fields appear only 
as $\Box$. The ghost action  will be analised later.
By combining (2.5-7) we get
$$
\li{ S^{(2)}_{tot} = & S^{(2)}+S_{GF}\cr
=&\int d^2x\sqrt g {\Big[} {1\over 4}\f h^{\m\n}\Box
h_{\m\n}-{1\over8}\f h\Box h+{3\over 8}D_{\r}\f hD^{\r}h\cr
&-{3\over 4}
D_{\b}\f h_{\r\m}D^{\b}h^{\r\m} 
-{1\over 2}D_{\n}\f D_{\r}h h^{\r\n}
+D_{\m}\f h^{\r\m}D^{\n}h_{\r\n}
+{1\over 2}\f h^{\m\l}h^{\n}_{\l}R_{\n\m}\cr
& +{1\over 8}\f Rh^2-{1\over 2}
   \f hh^{\m\n}R_{\m\n}+{1\over 2}h\hat \f R
  -{1\over 2}\hat \f\Box h-\hat \f h^{\m\n}R_{\m\n}+  \cr
&
  ({1\over 8}h^2-{1\over 4}h_{\m\n}h^{\m\n})V(\f )+{1\over 2}
  V^{\prime \prime}(\f)
\hat \f ^2+{1\over 2} V^{\prime}(\f )h\hat \f  \cr 
  &-{1\over 2}g^{\m\n}\pa _{\m}f\pa_{\n}f
+ \pa _{\m}f_0\pa _{\n}f(h^{\m\n}-{1\over 2}hg^{\m\n})
-{1\over 2}\pa  _{\m}f_0\pa _{\n}f_0(-{1\over 4}g^{\m\n}h_{\a\b}h^{\a\b}-{1\over 2}
hh^{\m\n}\cr
&+ h^{\m\l}h_{\l}^{\n}+{1\over 8}g^{\m\n}h^2)
 -{1\over 4}\f Rh_{\m\n}h^{\m\n}-{1\over 2}\f  h^{\n\b}h^{\m\a} R_{\a\n \b\m}-
 {1\over 2 \f}(\pa \hat \f)^2 \cr
&-{1\over 2}D_{\m}D_{\r}\f h^{\m\b}h^{\r}_{\b}+{1\over 2}D_{\m}D_{\n}\f 
   h^{\m\n}h{\Big]}. & (2.8)\cr }
$$ 
In order  to  remove $\f $ from the kinetic terms for $h$ in (2.8) we rescale
$$\sqrt \f h_{\m\n}\rightarrow h_{\m\n}\quad, \quad 
{\hat \f\over \sqrt\f}\rightarrow  \hat \f \quad. \eqno(2.9)$$
This field redefinition does not change the path-integral measure,
since the Jacobian of the transformation (2.9) is equal to one. This can be
shown by using the dimensional regularization and the 
delta-function identity $\d^{(2+\e)}(0)=0$.
$S^{(2)}_{tot}$ then changes as follows
$$\li{ S^{(2)}_{tot} = &\int d^2x\sqrt g {\Big[}{1\over 4} h^{\m\n}\Box 
h_{\m\n}-{1\over8} h\Box h-D_{\r}\F hD^{\r}h\cr
+&2
D_{\b}\F h^{\r\m}D^{\b}h_{\r\m} 
+ D_{\n}\F D_{\r}h h^{\r\n}\cr 
+& 
{1\over 2}h^{\m\l}h^{\n}_{\l}R_{\n\m} +{1\over 8}h^2R-
{1\over 4}Rh_{\m\n}h^{\m\n}\cr 
- &{1\over 2}
    hh^{\m\n}R_{\m\n}+{1\over 2}h\hat \f R
  -\hat \f h^{\m\n}R_{\m\n}+{1\over \f}
  ({1\over 8}h^2-{1\over 4}h_{\m\n}h^{\m\n})V(\f ) +{1\over 2}
  V^{\prime }(\f )\hat \f h\cr 
&+{1\over 2}V^{\prime\prime }(\f )\f \hat \f ^2-{1\over 2}g^{\m\n}\pa _{\m}f\pa_{\n}f
+ {1\over \sqrt \f} \pa _{\m}f_0\pa _{\n}f(h^{\m\n}-{1\over 2}hg^{\m\n})\cr
-&{1\over 2\f }\pa  _{\m}f_0\pa _{\n}f_0(-{1\over 4}g^{\m\n}h_{\a\b}h^{\a\b}-
{1\over 2}
hh^{\m\n}+
 h^{\m\l}h_{\l}^{\n}+{1\over 8}g^{\m\n}h^2)\cr
 -&{1\over 2}  h^{\n\b}h^{\m\a} R_{\a\n \b\m}
 +h_{\m\n}h^{\m\n} ({1\over 4}\Box \F+{7\over 4}(\nabla \F)^2)\cr
 +&h^2(-{1\over 8}\Box \F-{7\over 8}(\nabla \F)^2)-2D_{\l}\F h^{\l}_{\m}D_{\n}
  h^{\m\n}\cr
  -&2D_{\l}\F D_{\n}\F h^{\l}_{\m}h^{\m\n}+D_{\n}\F D_{\r}\F
hh^{\r\n}-{D^{\r}D_{\m}\f \over 2\f}h^{\m\b}h_{\r\b}+{ D_{\m}D_{\l}\f
\over 2\f }hh^{\l\m}\cr
+&{1\over 2}D_{\m}\hat  \f D^{\m}h+{1\over 2}D_{\m}\hat  \f  h D^{\m}\F-{1\over 2}
\hat \f D_{\m}\F D^{\m}h-{1\over 2}\hat \f h  (\nabla  \F)^2-{1\over 2}(\nabla
\hat \f)^2\cr
 &+\hat \f D_{\m}\hat \f D^{\m}\F -{1\over 2}\hat \f ^2(\nabla \F )^2 
 {\Big]}\quad, &(2.10)\cr }
$$ 
where $\f =e^{-2\F }$. Instead of using $h_{\m\n}$ it is more convinient
to use $\bar h_{\m\n}
=\P _{\m\n}^{\r\s} h_{\r\s} =h_{\m\n}-{1\over D} hg_{\m\n}$ and $ h $.
We take $D=2+\e$ as the dimension of the
spacetime, since we are going to use the dimensional regularization \cite{thv},
and 
$$ \P ^{\m\n}_{\r\s}={1\over 2} (\d^{\m}_{\r}\d^{\n}_{\s}+\d^{\m}_{\s}
 \d^{\n}_{\r}
)-{1\over D}g^{\m\n}g_{\r\s}\quad,$$
is the projector onto traceless states.
We will also use the doubling  trick of ref. \cite{thv}, so that we relabel the 
quantum fields $\{\bar h^{\m \n},\ h,\ \hat\f ,\ f \}$ in (2.10)
as  $\{\bar h^{'\m \n},\ h^{\prime},\ \hat\f ^{\prime},\ f^{\prime}\}$ and 
add to the action (2.10) the same action, but with
the quantum fields  relabeled as
$\{\bar h^{\prime \prime \m \n},\ h^{\prime \prime },\ 
\hat\f ^{\prime \prime},\ f^{\prime \prime }\}$. This allows us
to work with the complex fields
$$\bar h^{\m\n }={\bar h^{\prime \m\n}  +i\bar h^{\prime \prime \m\n}\over \sqrt 2},\   
\bar h^{*\m\n }={\bar h^{\prime \m\n}  -i\bar h^{\prime \prime \m\n}
\over \sqrt 2}, ... ,$$
so that (2.10) can be rewritten as
$$S^{(2)}_{tot}={1\over 2}\int dx \sqrt g 
(\bar h^{*\m\n}\ h^*\ \hat \f ^* \ f^*)
\hat K(I\Box +\hat K^{-1}\hat M)
 \left( \matrix {\bar h_{\r\s}\cr
h\cr \hat \f \cr f\cr }\right)\quad, \eqno(2.11)$$
where $I=diag (\P^{\m\n}_{\r\s},\ 1,\ 1,\ 1)$,
$$\hat K=  \left( 
\matrix {\P^{\r\s}_{\m\n}& 0&0&0\cr
0&-{\e \over 2(2+\e )} &-1&0\cr
0&-1 &2 &0\cr
0&0&0&2 \cr }\right),\ \eqno(2.12)
$$    
$$ \hat K^{-1}\hat M= \left( \matrix {\hat V^{\r\s}_{\a\b}&
\hat G_{\a\b}&\hat H_{\a\b}&\hat W_{\a\b}\cr
\hat M^{\r\s}&\hat P&\hat Q&\hat X\cr
\hat N^{\r\s}&\hat L&\hat S&\hat E \cr
\hat Y^{\r\s}&\hat Z&0&0 \cr }\right)\quad.\eqno(2.13)$$  
The  matrix elements in (2.13) are defined as
$$\li {\hat V_{\a\b}^{\r\s}=&-\P _{\a\b}^{\r\s}R+\P_{\a\b}^{\m\n}(R^{\r}_{\m}
\d ^{\s}_{\n}+R^{\s}_{\m}\d ^{\r}_{\n}-R^{\r\ \ \s}_{\ \m\n \  }
-R^{\r\ \ \s}_{\ \n\m\  })\cr
&-3\P^{\r\s}_{\a\b}\Box \F+7\P^{\r\s}_{\a\b}(\nabla
\F)^2+
2D_{\l} \F(\P^{\s a}_{\a\b}g^{\l\r}+\P^{\r a}_{\a\b}g^{\l\s}\cr
&-\P^{\l\s}
_{\a\b}g^{a\s}-\P^{\l\r}_{\a\b}g^{a\s}){\overrightarrow D}_a
+4\P ^{\m\n}_{\a\b}
(\d ^{\r}_{\m}D^{\s}D_{\n}\F\cr
 &-2\d ^{\r}_{\m}D^{\s} \F D_{\n}\F +\d
^{\s}_{\m} D^{\r}D_{\n}\F  -2\d ^{\s}_{\m}D^{\r}\F D_{\n}\F)+ O(e^{2\F }),
&(2.14)\cr }$$
$$\hat G_{\a\b}=\P^{\m\n}_{\a\b}\left({2-\e \over 2+\e } 
R_{\m\n}+2{3\e -2\over
2+\e }D_{\m}\F D_{\n}\F 
-2{\e  -2\over 2+\e }D_{\m}D_{\n}\F +2D_{\m} \F
 {\overrightarrow D}_{\n}\right), \eqno(2.15)$$    
$$\hat H_{\a\b}=-2\P ^{\m\n}_{\a\b}R_{\m\n},\eqno(2.16)$$
$$\hat W_{\a\b}=2\P ^{\m\n}_{\a\b}e^{\F }\pa _{\n}f_0
{\overrightarrow \pa}_{\m}\eqno(2.17)$$
$$\hat M^{\r\s}={2\e \over 1+\e }R^{\r\s}-{2+\e \over 1+\e }
\left(2{3\e -2\over 2+\e }D^{\r}\F D^{\s}\F 
-2 {\e -2\over \e +2}D^{\r}D^{\s}\F +(
\overleftarrow{\pa^\r}D^{\s}\F  +\overleftarrow{\pa^\s} D^{\r}\F )\right) ,
\eqno(2.18)$$
$$ \li {\hat P=&-{\e ^2\over (1+\e )(2+\e )}R -
{2+\e \over 1+\e }{\Big(}-D^{\m} \F \vec
\pa _{\m}\cr
&-{-\e ^2 +5\e +6\over (2+\e )^2 }\Box\F+{-4\e ^2 +3\e +6\over
(2+\e )^2}(\nabla \F )^2{\Big)}+ O(e^{2\F }), &(2.19)\cr }$$
$$\hat Q=-{\e \over1+\e  }R+{2+\e \over 1+\e }(2 \overleftarrow{\pa^\m} D_{\m}
\F +2\Box
 \F +2(\nabla \F )^2)+ O(e^{2\F }), \eqno(2.20)$$
 $$\hat X= -e^{\F }{\e \over 1+\e }\pa_{\m}f_0 {\overrightarrow \pa}^{\m},
 \eqno(2.21)$$
$$\li{\hat N^{\r\s}=&-{1 \over 1+\e }R^{\r\s}-{2+\e \over 2(1+\e )}{\Big(} 2
 {3\e -2\over 2+\e }D^{\r}\F D^{\s}\F 
-2 {\e -2\over \e +2}D^{\r}D^{\s}\F \cr
 &+(\overleftarrow{\pa^\r}D^{\s}\F  + \overleftarrow{\pa^\s} D^{\r}\F ){\Big)}
\quad,&(2.22)\cr}$$
$$\hat L={\e \over2(2+\e  )(1+\e )}R  -{\e \over 2(1+\e )}D^{\m}\F 
{\overrightarrow \pa}_{\m}
+{2-\e \over 2+\e }\Box \F -{-3\e ^2+6\e +8\over 2(1+\e )(2+\e )}(\nabla \F )^2
+ O(e^{2\F }) , \eqno(2.23)$$
$$\hat S= -{\e \over 2(1+\e )}R+{2+\e \over 1+\e }\overleftarrow{\pa^\m}D_{\m}\F
+{1\over
1+\e }\Box \F  +{1\over 1+\e  }(\nabla \F )^2 + O(e^{2\F }), \eqno(2.24)$$
$$\hat E={\e \over 2(1+\e )}e^{\F }\pa ^{\m}f_0\vec \pa _{\m},
\eqno(2.25)$$
$$\hat Y^{\r\s}=\overleftarrow{\pa^\r}e^{\F }\pa^{\s}f_0,\eqno(2.26)$$
$$ \hat Z=-{\e  \over 2(2+\e )}\overleftarrow{\pa^\r} e^{\F }\pa _{\r} f_0 \quad.
\eqno(2.27)$$

In the lowest-order weak-coupling approximation the matrix elements of 
(2.13) do not depend on the potential $V(\f )$, 
because the terms
${V(\f )\over \f },\ V^{\prime}(\f ),\ V^{\prime\prime}
(\f )\f $ are of $O(\f^{-k-1})$, and we can discard them . Therefore
the leading-order weak-coupling approximation will not depend on the
potential $V$.   

After fixing of the quantum gauge symetries, we must introduce the 
corresponding ghost fields.
 Under the general coordinate transformation the quantum metric
$h_{\m\n} $ and the field $\hat \f  $ transform as follows
$$\d _0h_{\m\n}=-D_{\m}\e _{\n}-D_{\n}\e _{\m}+(D_{\n}h^{\r}_{\m}+
D_{\m}h^{\r}_{\n} -D^{\r}h_{\m\n})\e _{\r}\quad, $$ 
$$ \d _0\hat \f =-\e ^{\r }\pa  _{\r}(\f +\hat \f )\quad.\eqno(2.28)$$
The ghost action is then given by
$$S_{gh}=\int d^2x\ \sqrt g \f\bar c^{\m}\ [-\Box c_{\m}-R^{\n
}_{\m}c_{\n}+{1\over \f }D_{\m}(c^{\r}\pa _{\r}\f )]\quad, \eqno(2.29)$$
where we write only the part of the ghost action which gives the contribution
to the one-loop effective action.
We can rescale 
$$\f \bar c^{\m}\rightarrow \bar c^{\m}\quad,\eqno(2.30)$$
so that
$$S_{gh}=\int d^2x \sqrt g \bar c^{\m} 
[-\Box c_{\m}-R^{\n}_{\m}c_{\n}+{1\over \f }D_{\m}(c^{\r}\pa _{\r}\f )] 
\quad. \eqno(2.31)$$

\sect{3. Expansion around a flat metric }

The calculation of the one-loop effective action can be simplified  
by expanding the
background metric around a flat metric $\h_{\m\n}$ as
$$g_{\m\n} = \h_{\m\n} +\g_{\m\n} + O(\g^2 ) \quad. \eqno(3.1)$$
After inserting (3.1) into (2.12) and (2.13), we get
$$\sqrt g (I\Box +\hat K^{-1}\hat M)=diag (P^{\r\s}_{\a\b},\, 1,\, 1,\, 1)\pa^2
 +  K^{-1} M \, ,\eqno(3.2)$$ 
where $\pa ^2 =\h ^{ab}\pa _a\pa _b$, $P^{\r\s}_{\a\b}$ is given in (3.8)
and
$$ K^{-1} M = \left( \matrix {{\tilde V}^{\r\s}_{\a\b}&G_{\a\b}&H_{\a\b}&W_{\a\b}\cr
M^{\r\s}&P&Q&X\cr
N^{\r\s}&L&S&E \cr
Y^{\r\s}&Z&0&F \cr }\right)\quad .\eqno(3.3)$$
The matrix  elements  in (3.3) which are relevant for our calculation
are given by
$$\li {{\tilde V}_{\a\b}^{\r\s}=&\overleftarrow{\pa_a}A^{ ab\r\s}_{\a\b}{\overrightarrow \pa}_{b}\cr
-&\left(\h ^{ab}
P^{\r\s}_{\e \t }S^{\e \t}_{b\a\b}
-2\sqrt g \pa _{\l}\F (g^{\l\r}\P ^{\s a}_{\a\b}+
\P ^{\r a}_{\a\b}g^{\l\s}-\P^{\l\s}_{\a\b}g^{a\r}-\P ^{\l\r}_{\a\b}g^{a\s} )
\right) {\overrightarrow \pa}_a  \cr
+&\overleftarrow{\pa_a} \h^{ab}P^{\e\t}_{\a\b}S^{\r\s}_{b\t\e }-\sqrt g \P _{\a\b}^{\r\s}R+\sqrt g \P _{\a\b}^{\m\n}(R^{\r}_{\m}
\d ^{\s}_{\n}+R^{\s}_{\m}\d ^{\r}_{\n}-R^{\r\ \ \s}_{\ \m\n \  }
-R^{\r\ \ \s}_{\ \n\m\  })\cr
-&3\sqrt g \P^{\r\s}_{\a\b}\Box \F +7\sqrt g \P^{\r\s}_{\a\b}(\nabla
 \F )^2+\cr 
+&4\sqrt g \P ^{\m\n}_{\a\b}
\left( \d ^{\r}_{\m}D^{\s}D_{\n}\F -2\d ^{\r}_{\m}D^{\s} \F D_{\n}\F +\d
 ^{\s}_{\m} D^{\r}D_{\n}\F  -2\d ^{\s}_{\m}D^{\r}\F D_{\n}\F \right)  \cr
-&2\sqrt g \pa_{\l}\F (\P^{a\t}_{\a\b}g^{\l\e}+\P^{\e a}_{\a\b}g^{\l
\t}\cr 
-&\P^{\l\e}_{\a\b}g^{a\t}-\P^{\l\t}_{\a\b}g^{a\e})S^{\r\s}_{a\e\t} +\h
^{ab}P^{\e\t}_{\d\h}S^{\d\h}_{a\a\b}S^{\r\s}_{b\e\t}+ O(e^{2\F})
\quad,&(3.4)\cr }$$
$$\li{P=&\overleftarrow{\pa_a}\bar \g^{ab}{\overrightarrow \pa}_b 
-{\e ^2\over (1+\e )(2+\e )}\sqrt g R 
-{2+\e \over 1+\e } \sqrt g{\Big(} -D^{\m} \F  
{\overrightarrow \pa}_{\m}\cr 
 &- {-\e ^2 +5\e +6\over (2+\e )^2 }\Box \F +{-4\e ^2 +3\e +6\over
(2+\e )^2}(\nabla \F )^2{\Big)} + O(e^{2\F })\quad,  &(3.5)\cr }$$
$$\li{S=& {\overleftarrow \pa}_a\bar \g^{ab} {\overrightarrow \pa}_b 
-{\e \over 2(1+\e )}\sqrt g R\cr 
&+\sqrt g {1\over 1+\e }\left( (2+\e ) \overleftarrow{\pa^\m} D_{\m}\F 
+\Box \F  +(\nabla \F )^2\right) +O(e^{2\F})\quad, & (3.6)\cr }$$
$$F={\overleftarrow \pa}_a \bar \g^{ab} {\overrightarrow \pa}_b \quad,
\eqno(3.7)$$
where
$$ P^{\m\n}_{\r\s}={1\over 2} (\d^{\m}_{\r}\d^{\n}_{\s}+\d^{\m}_{\s}
\d^{\n}_{\r}
)-{1\over D}\h^{\m\n}\h_{\r\s},$$
$$S^{\r\s}_{a\m\n}=2\G^{(\r}_{a(\m}\d^{\s)}_{\n)},$$
$$\bar \g^{\m\n} = \g^{\m\n}-{1\over 2}\g \h^{\m\n},$$
$$A^{\r\s ab}_{\a\b}=P^{\r\s}_{\a\b}\bar \g^{ab}-{1\over D}\h^{ab}
(\g^{\r\s}\h_{\a\b}-\g_{\a\b}\h^{\r\s})
\quad, \eqno(3.8)$$
and $\g =\g^{\m\n}\h_{\m\n}$. 
In the case of the ghost action (2.31),  expansion (3.1) yields
$$\li{S_{gh}=&\int d^2x\ \bar c^{\m} (\d^{\n}_{\m}\pa^2 +T^{\n}_{\m})c_{\n}\cr
=&\int dx\bar c^{\m}{\Big[} \d^{\n}_{\m}\pa^2+\d^{\n}_{\m}{\overleftarrow \pa}_a
{\bar \g}^{ab}
{\overrightarrow \pa}_{b}-\G^{\n}_{a\m} \h^{a\s} {\overrightarrow\pa}_{\s} +
\h^{ab}\G^{\r}_{a\m}
\G^{\n}_{b\r}\cr 
&+{\overleftarrow\pa}_{\s}\h ^{a\s}\G
^{\n}_{a\m}+2\pa _{\r}\F (\h ^{\r\n}-{\bar \g}^{\r\n}){\overrightarrow \pa}_{\m}
 -2\h ^{\r\a}\G ^{\n}_{\a\m}\pa _{\r}\F +\sqrt g R_{\m}^{\n}\cr
 &-2\d ^{\r\n}\G ^{\a}_{\r\m}\pa _{\a} \F - (\h ^{\r\n}-\bar \g
^{\r\n})( 4\pa _{\m}\F \pa _{\r}\F  -2\pa _{\m}\pa _{\r}\F ){\Big]} c_{\n}
\quad.&(3.9)\cr }$$
After a suitable transformation of the complex fields in the 
path-integral (2.2), we
get from (2.3) the following one-loop contribution the effective action 
$$\G _1={i\over 2}{\rm Tr} \log \left( 1+K^{-1}M{1\over  \pa^2}\right)
-i{\rm Tr} \log \left( 1+T{1\over \pa^2}\right), \eqno(3.10)$$
where $K^{-1}M$ and $T$ are defined by (3.3) and (3.9).   
After expanding the logarithm in (3.10), we obtain
$$\li{\G_1 =&{i\over 2} {\Big [}({\tilde V}^{\r\s}_{\a\b}P^{\a\b}_{\r\s}+P+S+F)
{1\over \pa^2}-{1\over 2}
({\tilde V}^{\r\s}_{\a\b}{1\over \pa^2}P^{\a\b}_{\m\n}{\tilde V}^{\m\n}_{\g\d}
{1\over \pa^2}P^{\g\d}_{\r\s}\cr
&+2G_{\a\b}{1\over \pa^2}P^{\a\b}_{\m\n}M^{\m\n}{1\over \pa^2}
+2H_{\a\b}{1\over \pa^2}P^{\a\b}_{\m\n}N^{\m\n}{1\over \pa^2}+\cr
&+2Y_{\a\b}{1\over \pa^2}P^{\a\b}_{\m\n}W^{\m\n}{1\over \pa^2}+
P{1\over \pa ^2}P{1\over \pa ^2}+
S{1\over \pa ^2}S{1\over \pa ^2}\cr
 &+F{1\over \pa ^2}F{1\over \pa ^2}
+2Q{1\over \pa ^2}L{1\over \pa ^2}+
2X{1\over \pa ^2}Z{1\over \pa ^2}){\Big]}\cr
&-i{\Big[}T^{\m}_{\n}\d ^{\n}_{\m}{1\over  \pa ^2}-{1\over 2}T^{\m}_{\n}\d ^{\n}_{\r}
{1\over  \pa ^2} 
T^{\r}_{\s}\d ^{\s}_{\m}{1\over  \pa^2}{\Big]}.\ &(3.11)\cr }$$

\sect{4. Calculation of the effective action for $\f$ =const.}

In this section we will compute the part of the effective action which
is independent of $\pa\f$, which can be done by taking $\f$ to be a spacetime
constant. Also, we will discard the terms proportional to $1/\f = e^{2\F}$, 
since they
give a sub-leading contribution in the weak-coupling approximation.  
First, we compute
the terms in (3.11) which contain $\tilde V$. 
We denote the vertices with two, one and zero spacetime
derivatives as $A$, $B$ and $C$, respectively.
The contribution due to $\bar h_{\m\n}$ loops can be written as
$$\li{ {\tilde V}{1\over \pa ^2}-{1\over 2}
{\tilde V}{1\over \pa ^2}{\tilde V}{1\over \pa ^2}
=& (A+B+C){1\over \pa ^2}\cr
&-{1\over 2} \left( A{1\over \pa ^2}A{1\over \pa ^2}+
B{1\over \pa ^2}B{1\over \pa ^2}+C{1\over \pa ^2}C{1\over \pa ^2} \right)\cr
 &-A{1\over \pa ^2}B{1\over \pa ^2}- A{1\over \pa ^2}C{1\over \pa ^2}
- B{1\over \pa ^2}C{1\over \pa ^2} \quad.&(4.1)\cr }$$
Note that
if we set $\F= const.$ in (3.4) and if we are not carefull
when calculating the corresponding one-loop 
contribution to the effective action, we may obtain a  result 
independent of $\F $, which is wrong. The term which requires a
carefull consideration is $\int dx \sqrt gR{1\over \Box  }\Box \F $, and it gives
a nonzero result for $\F = const.$, which is
$\int dx\sqrt gR\F$.

Now, one can show that $A$ and $B$ diagrams are equal zero 
(see the Appendix). The diagram $C$ is given by
(formula (A.5) of the Appendix) 
$$\li {C =& -{i\p ^{{D\over 2}}\over (2\p )^2}\G (-{\e \over 2})
\int dx\ \left[ \left({-D^2+D+2\over 2}
 -{4\over D}\right)\sqrt g R+\h ^{ab}P^{\e\t}_{\d\h}P^{\a\b}_{\r\s}S^{\d\h}_{a\a\b}
S^{\r\s}_{b\e \t}\right]  \cr 
&+4{i\p^{{D\over 2}}\over (2\p)^2}\G (-{\e \over 2}){D+2\over 2}\int
dx \sqrt g R\F .&(4.2)\cr }$$
Note that the last term in the first line of (4.2) is a non-covariant quantity.
We will denote the diagram which contains a vertex $X$ and a vertex $Y$ as
$XY$, where $X,Y \in \{ A,B,C, ...\}$. The 
$AA$ diagram is given by 
$$AA = \int dxdy A^{\r\s ab}_{\a\b}(x)A^{\m\n cd}_{\g\d}(y)
P^{\a\b}_{\m\n}P^{\g\d}_{\r\s}\pa _{a}^{x} \pa _{d}^{y} G(x-y)\pa _b^x
\pa _c^yG(y-x)\quad,\eqno(4.3)$$ 
where the Green's function  $G(x-y)$ satisfies
$$\Box _xG(x-y)=\d (x-y) \quad.\eqno(4.4)$$
By using (A.8), (4.3) becomes
$$- {i\pi ^{{D\over 2}}\over (2\p )^2} {D^2+D-2\over 2}\G (1-{\e \over 2})
B(2+{\e \over 2}, 2+{\e \over 2})\left( \int dx\sqrt gR{1\over \Box }R+
{4\over \e(1+{\e \over 2})}
\int dx\sqrt g R\right)  , \eqno(4.5)$$
where 
$$\int dx\sqrt gR{1\over \Box }R=\int dxdy \sqrt {g(x)}\sqrt {g(y)}
R(x)G(x-y)R(y).\eqno(4.6)$$
The $AB$ diagram is equal to zero. 
The $AC$ diagram is given by
 $$\int dxdy A^{\r\s ab}_{\a\b}(x) C^{\m\n}_{\g\d}(y)\pa _b^xG(x-y)\pa _a^xG(y-x) 
P^{\a\b}_{\m\n}P^{\g\d}_{\r\s}.\eqno(4.7)$$
If we use (A.10) and 
$$\li{A^{\r\s
ab}_{\a\b}C^{\m\n}_{\g\d}P^{\a\b}_{\m\n}P^{\g\d}_{\r\s} 
=&\left( -{D^2+D
-2\over 2 }+{D^2-4\over D}\right) {\bar\g}^{ab} R \cr
&+\left({4\over D}-{3\over 2}\right)(D^2+D-2)\bar \g^{ab}
\Box \F ,&(4.8)\cr}$$ 
where $R=\pa_a\pa_b \g^{ab}-\Box \g + O(\g ^2)$,
we get 
$$AC=2{i\p ^{{D\over 2}}\over (2\p )^2} \int dx\sqrt g\left( 
R{1\over \Box }R-R\F \right). \eqno(4.9)$$
The $BB$ diagram is given by
 $$\li {BB =\int & dxdy {\Big[} (-\h
^{ab}P^{\r\s}_{\e\t}S^{\e\t}_{b\a\b}(x)+ 4D_{\l} \F (\P ^{a\s}_{\a\b}g^{\l\r}
-\P^{\l\s}_{\a\b} g^{a\r}))
 {\overrightarrow \pa}_a\cr 
 & +\overleftarrow{\pa^a} P^{\e\t}_{\a\b}S^{\r\s}_{a\e\t}(x){\Big]} G(x-y)
P^{\a\b}_{\m\n}P_{\r\s}^{\g\d}{\Big[}(
-P^{\m\n}_{\c\h}S^{\c\h}_{c\g\d}(y)\h  ^{cd}\cr 
&+ 4D_{\k}\F (y)(\P^{\n d}_{\g\d}g^{\m\k}-\P^{\n\k}_{\g\d}g^{d\m}))
\overrightarrow\pa_d + 
\overleftarrow{\pa^c} P^{\h\c}_{\g\d}S^{\m\n}_{c\h\c}(y){\Big]}
  G(y-x). &(4.10)\cr }$$
From (A.13) and (4.10) we get   
$$\li {BB=&-4{i\p^{{D\over 2}}\over (2\p )^2}
B(1+\e/2,1+\e/2)\cr
\cdot &{\Big[}\G (1-\e/2)\int dxdyG(x-y)P^{\a\b}_{\g\d}
P^{\m\n}_{\r\s}\pa^a
S^{\g\d}_{a\m\n}(x)\pa ^cS_{c\a\b}^{\r\s}(y)\cr 
&+ {1\over 2}\G (-{\e \over 2})(\h^{ac}\int
 dx
P^{\a\b}_{\g\d}P^{\m\n}_{\r\s}
S^{\g\d}_{c\m\n}(x)S_{a\a\b}^{\r\s}(x)\cr 
&-4\int dx S^{\e\t}_{a\a\b}P^{\a\b}_{\m\n}
(P^{a\m}_{\e\t}g^{\l\n}-P^{\l\m}_{\e\t}g^{a\n})\pa _{\l}\F ){\Big]}, &(4.11)\cr }$$
were we have taken into account that $\F$ is a spacetime constant. Since  
$$B(1+\e/2,1+\e/2)=1-\e +O(\e^2),\eqno(4.12)$$
(4.11) can be rewritten as 
$$\li {BB=&-4{i\p ^{{D\over 2}}\over (2\p )^2}
{\Big \{} \G (1-{\e \over 2}){\Big[}\int dxdyG(x-y)P^{\a\b}_{\g\d}
P^{\m\n}_{\r\s}\pa^a
S^{\g\d}_{a\m\n}(x)\pa^cS_{c\a\b}^{\r\s}(y) \cr 
&+\h^{ac}\int dx
P^{\a\b}_{\g\d}P^{\m\n}_{\r\s}
S^{\g\d}_{c\m\n}(x)S_{a\a\b}^{\r\s}(x){\Big]}+{1\over 2}\G (-\e/2)
\int dx (\h^{ac}
 P^{\a\b}_{\g\d}P^{\m\n}_{\r\s}
S^{\g\d}_{c\m\n}S_{a\a\b}^{\r\s}\cr 
& -4(1-\e ) S^{\e\t}_{a\a\b}P^{\a\b}_{\m\n}
(g^{\l\n}P^{a\m}_{\e\t}-P^{\l\m}_{\e\t}g^{a\n})\pa_{\l}\F ){\Big \}}
. &(4.13)\cr }$$
Covariantization of (4.13) gives
$$\li {BB=&-8{i\p ^{{D\over 2}} \over (2\p )^2}\left( \int dx \sqrt g
R{1\over \Box }R+4\int dx\sqrt gR\right) \cr
-2&{i\p ^{{D\over 2}}\over (2\p )^2 }\G (-{\e \over 2})\int dx  
[\h ^{ab}P^{\e\t}_{\d \h}P^{\a\b}_{\r\s}S^{\d\h}_{a\a\b}
S^{\r\s}_{b\e \t}-2(1-\e )(D+2)\sqrt g R\F ]. &(4.14)\cr }
$$
By summing up the non-covariant terms in (4.14) and (4.2) we obtain 
a general coordinate invariant result. $BC$ and $CC$ diagrams are
equal to zero. 

There is also a contributions to the effective action from the term 
$P^2+S^2+NF^2$. By using (A.8) we get
$$\li {P^2+S^2+NF^2 =&- (N+2){i\p^{{D\over 2}}\over (2\p )^2}\G (1-{\e \over
2})B(2+{\e \over 2},2+{\e \over 2})
{\Big(} \int dx\sqrt g  R{1\over \Box }R \cr 
&+{4\over \e(1+{\e \over 2})}\int dx\sqrt gR {\Big)}
 -2{i\p^{{D\over 2}}\over (2\p )^2}
\int dx \sqrt g R\F \quad.&(4.15)\cr }$$
The other terms in (3.11) are either zero or of a 
sub-leading order in $e^{2\F }$.
   
The contribution due to the ghost loops is given by
the last line in (3.11)
$$\li { T{1\over  \pa^2}-{1\over 2}  T
{1\over  \pa^2} T{1\over  \pa^2} =&
 (\bar A+\bar B+\bar C){1\over \pa ^2}\cr
-&{1\over 2}\left(\bar A{1\over \pa ^2}\bar A{1\over \pa ^2}+
\bar B{1\over \pa ^2}\bar B{1\over \pa ^2}+\bar C{1\over \pa ^2}\bar C
{1\over \pa^2} \right)\cr
-&\bar A{1\over \pa ^2}\bar B{1\over \pa ^2}-\bar A{1\over \pa ^2}\bar C
{1\over \pa^2}- \bar B{1\over \pa ^2}\bar C{1\over \pa ^2}
\quad.&(4.16)\cr }$$
Again, we calculate only the diagrams $\bar A,\bar B,\bar C,\bar A\bar A,
 \bar B\bar B,\bar C\bar C,\bar A\bar B,\bar A\bar C,\bar B\bar C$,
 which apper in (4.16). The calculation is similar to 
 the previous calculation, so that we give only the list of
results
$$\bar A=\bar B=\bar A\bar B=\bar B\bar C=0,\eqno(4.17) $$
$$\li {\bar C&= \int dx  (\sqrt gR+\h
^{ab}\G _{a\m}^{\r}\G ^{b}_{\n\r}-\pa _{\m}\F \pa ^{\m}\g )G (0)\cr 
&=-{i\p ^{{D\over 2}}\over (2\p )^2}\G (-{\e \over 2})\int dx\left( \sqrt g R+
\h
^{ab}\G  _{a\n}^{\r}\G _{b\r}^{\n}+\F \Box \g \right) ,&(4.18)\cr }$$
$$\li {&\bar A\bar A=D\int dxdy\bar \g ^{ab}(x)\bar \g ^{cd}(y)
 \pa_a^x\pa _{d}^{y} G(y-x)\pa_b^x
\pa_c^yG(x-y)\cr
=&- {i\pi^{{D\over 2}}\over (2\p )^2} D\G (1-{\e \over 2})
B(2+{\e \over 2}, 2+{\e \over 2})\left(\int R{1\over \Box }R+{4\over \e(1+{\e \over 2})}
\int dx\sqrt g R\right), & (4.19)\cr }$$
$$\bar A\bar B={i\p^{{D\over 2}}\over (2\p )^2}\int dx \sqrt g R\F ,
 \eqno(4.20)$$
$$\bar A\bar C =-{i\p^{{D\over 2}}\over (2\p )^2}
\int dx \sqrt g \left( R{1\over \Box }R+2R\F \right),\eqno(4.21)$$
$$\bar B\bar B=-2{i\p^{{D\over 2}}\over (2\p )^2}\left( \int
dx\sqrt g(R{1\over \Box }R+4 R-2R\F )+\G (-{\e \over 2})\int dx
 (\G^{\n}_{c\m}\G^{\m}_{d\n}\h^{cd}+\F \Box \g )\right) .\eqno(4.22)$$
By collecting (4.1-2), (4.5), (4.9), (4.14-15) and (4.17-22), we get that
the $\pa\f$-independent part of $\G_1$ is given by
$$ \G _1^{(1)}=-{N-24\over 96\p }\int dx \sqrt gR{1\over \Box }R -{N-24\over 24
\p \e}\int dx \sqrt g R\ +{5\over 4\p }\int dx \sqrt g R\F .
\eqno(4.23)$$

\sect{5. Calculation of the effective action for $g_{\m\n}=\h_{\m\n}$}

In this section we will compute $\G_1$ in the special case
of the flat background, i.e. when $g_{\m\n}=\h_{\m\n}$.
We start from (3.11) and by using (3.4) we get
$$\li  {{\tilde V}^{\r\s}_{\a\b}=& 2D_{\l}\F (\h^{\r\l}P^{a\s}_{\a\b}+
\h ^{\l\s}P^{a\r}_{\a\b}
-\h^{a\s}P^{\l\r}_{\a\b}-P^{\l\s}_{\a\b} \h^{a\r}){\overrightarrow \pa}_a\cr
  -&3P^{\r\s}_{\a\b}\Box \F +7P^{\r\s}_{\a\b}(\nabla
 \F )^2 \cr
+&4 P ^{\m\n}_{\a\b}
(\d ^{\r}_{\m}D^{\s}D_{\n}\F -2\d ^{\r}_{\m}D^{\s} \F D_{\n}\F +\d
 ^{\s}_{\m} D^{\r}D_{\n}\F  -2\d ^{\s}_{\m}D^{\r}\F D_{\n}\F )\cr
 +&O(e^{2\F }). &(5.1) \cr}$$
Since the terms with two derivatives do not appear in (5.1), our calculation
will be simpler. Again,
we will denote the terms with one derivate as $B$, and the terms without 
derivates as $C$.  
By using (A.5) we get    
$$C=-{i\p^{{D\over  2}}\over (2\p )^2}(D^2+D-2)\G(-\e/2) \left({7\over 2}
-{8\over D}\right)\int dx (\nabla \F )^2 \quad.\eqno(5.2)
$$

The BB  diagram is given by
$$\li {BB=&4\int dxdy 
D_{\l}\F (x)(\h^{\r\l}P^{a\s}_{\a\b}+\h^{\s\l} P^{a\r}_{\a\b}-
\h^{a\s}P^{\l\r}_{\a\b}-P^{\l\s}_{\a\b}\h^{a\r})\overrightarrow{\pa_a^x} 
G(x-y)\cr  
&D_{\k}\F (y)(\h ^{\m\k}P^{b\n}_{\g\d}+\h ^{\n\k}P^{b\m}_{\g\d}-
\h^{b\n}P^{\k\m}_{\g\d}-P^{\k\n}_{\g\d}\h^{b\m})\overrightarrow{\pa_b^y} 
G(y-x)P^{\a\b}_{\m\n}P_{\r\s}^{\g\d}. &(5.3)\cr }$$
By using (A.13) we obtain
$$BB=4{i\p^{{D\over 2}}\over(2\p )^2}\G (-{\e \over 2})B(1+{\e
\over 2},1+{\e \over 2})(D^2+D-2)\int dx (\nabla \F )^2.\eqno(5.4)$$

The matrix elements $P$ and $S$ are given by 
the expresssions (3.5) and (3.6). After inserting $g_{\m\n}=\h _{\m\n}$ we get
$${1\over 2}\left(  P{1\over \pa ^2}-{1\over 2}P{1\over \pa ^2}P{1\over
\pa ^2}\right) =-{1\over  2}{i\p ^{{D\over 2}}\over  (2\p )^2} 
\G(-{\e \over 2})
\left( {4\e ^2-3\e-6\over  (1+\e  )(2+\e )}
 -{1\over 4}\left({2+\e \over 1+\e }\right)^2 \right)\int dx(\nabla  \F)^2
, \eqno(5.5)$$
and similarly
$${1\over  2}(S{1\over \pa ^2}-{1\over 2}S{1\over \pa ^2}S{1\over
\pa ^2}) =-{1\over  2}{i\p ^{{D\over 2}}\over  (2\p )^2}\G (-{\e \over2}) 
\left({1\over 1+\e }
-{1\over 4}\left( {2+\e \over 1+\e }\right)^2\right)
\int dx(\nabla \F )^2
\quad.\eqno(5.6)$$

Next,  we will calculate the $GM$  diagram. From (2.15) and (2.18) we get
$$\li {GM&=\int dxdyG_{\a\b}(x) G(x-y)M^{\r\s}(y)P^{\a\b}_{\r\s}G(x-y)  \cr
&=-4{2+\e \over 1+\e }\int dxdy P^{\m\n}_{\r\s}D_{\m}\F (x)D^{\s}\F
(y)\pa _{\n}^x\pa _y^{\r}G(x-y)G(y-x),&(5.7)\cr }$$  
where we have descarded the terms which vanish after the integration. 
By using (A.13) we get
$$\li{GM=&-4{i\p^{{D\over 2}}\over (2\p )^2}{2+\e \over 1+\e }
\left({1\over 2}
\G (-{\e  \over 2})B(1+{\e \over 2},1+{\e \over 2}){D^2+D-2\over 2D}
-\G (1-\e/ 2)(1-1/D)\right)\cr
&\cdot\int dx (\nabla\F )^2 \quad. &(5.8)}$$

It is easy to see from  (2.16) and ( 2.22) that
the diagram $HN$ vanishes.

The diagram $QL$ is given by 
$$QL=-\e {2+\e \over (1+\e )^2}\int dxdy D_{\m}\F (x) D_{\n} \F (y)
G(x-y)\pa ^{\m}_x\pa ^{\n}_yG(y-x).\eqno(5.9)$$
From (A.13) we  get
$$QL=2{i\p ^{{D\over  2}}\over (2\p )^2}\int dx
(\nabla \F)^2 \quad.\eqno(5.10)$$

The diagrams $WY$ and $XZ$ are of the order $e^{2\F }$, and we
will descard them in the weak-coupling approximation.

The contribution from the ghost loops is determined by
$$T_{\m}^{\n}=2\pa^{\n}\F {\overrightarrow \pa}_{\m} -4\pa_{\m}\F \pa^{\n}\F +
2\pa_{\m}\pa^{\n}\F, \eqno(5.11)$$
which is  obtained from (3.9) by taking $g_{\m\n}=\h_{\m\n}$.
(5.11) gives only two  diagrams differant from zero. These are
$\bar C$ and $\bar B \bar B$. From (A.5) we get
$$\bar C={4i\p ^{{D\over 2}}\over (2\p )^2 }\G (-{\e \over 2})\int dx 
(\nabla \F )^2 \quad,\eqno(5.12)$$
while (A.13) gives
$$\bar B \bar B= -2{i\p ^{{D\over 2}}\over (2\p )^2}\G (-{\e \over
2})\int (\nabla \F )^2.\eqno(5.13)$$
By colecting  (3.11), (5.2), (5.4-6), (5.8), (5.10), (5.12-13) we get
that the one loop correction to the
effective action in the case of the flat background metric is given  by
$$ \G^{(2)}_1=-{\p^{{D\over 2}}\over 2(2\p )^2}{\Big[} -8\G (-\e/2)
+23{\Big]}\int dx (\nabla \F )^2 \quad.\eqno(5.14)$$

\sect{6. The complete one-loop effective action}

In sections 4 and 5 we have found the one-loop effective action in two
special  cases: $\f =const.$ and $g_{\m\n}=\h _{\m\n}$.  In this
section we will compute the complete one-loop effective action by
adding to $\G_1\sp{(1)} + \G_1\sp{(2)}$ the terms which vanish
for $\f =const.$ and $g_{\m\n}=\h _{\m\n}$. 
It is easy to see from (3.4-7) and (2.14-27), that there
is only one such term in the weak-coupling approximation. It is given by
$$\int dxdy\sqrt {g(x)} \sqrt {g(y)} R(x)G(x-y)(\nabla \F (y))^2\ ,\eqno(6.1)$$
and our task is to determine the coefficient which multiplies it.

We start from the diagrams in (4.1). The term (6.1) appears only in
the diagram $AC$, which is given by
$$\li{AC =\int dxdy &[P^{\r\s}_{\a\b}\bar \g ^{ab}(x)-{1\over D}(\g^{\r\s}(x)
\h_{\a\b}-\g _{\a\b}(x)\h ^{\r\s})][7P^{\m\n}_{\g\d}(\nabla
\F)^2\cr 
&-16P^{\m\e}_{\g\d}\pa ^{\n}\F (y)\pa _{\e}\F (y)]\pa _b^xG(x-y)\pa
_a^xG(y-x)P^{\a\b}_{\m\n}P^{\g\d}_{\r\s}.&(6.2)\cr }$$
In (6.2) we have written only the relevant terms of the C vertex. 
From (A. 10) we get
$$AC = 2 {i\p ^{{D\over 2}}\over (2\p )^2}\int dxdy (\pa _a\pa _b\bar \g
^{ab}(x) -{1\over 2} \Box \g (x)) (\nabla \F (y))^2 G(x-y)+O(\g ^2).\eqno(6.3)$$
By using $R= \pa _a\pa _b\bar \g^{ab} -{1\over 2} \Box \g  +O(\g ^2)$ 
we can rewritte (6.3) in the form
$$AC =2 {i\p ^{{D\over 2}}\over (2\p )^2}\int dxdy \sqrt {g(x)}\sqrt {g(y)}R(x)
 (\nabla \F (y))^2 G(x-y).\eqno(6.4)$$  

The term (6.1) appears in $PP$ and $SS$ diagrams,
whose contribution is
$$2 {i\p ^{{D\over 2}}\over (2\p )^2}\int dxdy \sqrt {g(x)}\sqrt {g(y)}R(x)
 (\nabla \F (y))^2 G(x-y)\quad.\eqno(6.5)$$  

(6.1) also appears in the diagrams with ghost loops. From (3.9) we see
that the relevant part of the $\bar A \bar C$ digram is 
$$\bar A \bar C= -4\int dxdy \bar \g ^{ab}(x)  \pa _{\m }\F (y)\pa
^{\m}\F (y)\pa _b^xG(x-y)\pa _a^xG(y-x).\eqno(6.6)$$  
By using (A.10) we get 
$$\bar A \bar C =4 {i\p ^{{D\over 2}}\over (2\p )^2}\int dxdy \sqrt {g(x)}
\sqrt {g(y)}R(x) (\nabla \F (y))^2 G(x-y).\eqno(6.7)$$  

After summing up (6.4), (6.5) and (6.7) we obtain 
$$\G _1^{(3)}= -{1\over 2\p }\int dxdy \sqrt {g(x)}\sqrt{g(y)}R(x)
 (\nabla \F (y))^2 G(x-y).\eqno(6.8)$$ 
By collecting (4.23), (5.14) and (6.8) we get the bare effective action
$$\li {{\bar \G}_1&=S-{N-24\over 96\p }\int dx \sqrt g R{1\over \Box }R 
-{N-24\over 24\p \e}\int dx \sqrt g R - {1\over 2\p }\int dx R
{1\over \Box } (\nabla \F )^2 \cr 
&-{\p ^{{\e \over 2}}\over 8\p }[ -8\G (-\e/2)+23]\int dx \sqrt g  
(\nabla \F )^2 
+{5\over 4\p }\int dx \sqrt g R\F  + O(e\sp{2\F}),&(6.9)
\cr }$$
where $S$ is the classical action (1.2).
After making
a modified minimal subtraction of the poles in (6.9) we get the renormalized 
one-loop effective action   
$$\li{\G_1 =&S-{N-24\over 96\p }\int dx \sqrt g R{1\over \Box }R
-{1\over \p }
\int dx \sqrt g \left( {1\over 2}R{1\over \Box  }
 (\nabla \F )^2 -{5\over 4}R\F +{23\over 8}(\nabla \F )^2\right)\cr
& + O(e^{2\F }) \quad.&(6.10)}$$  
The expression (6.10) is our final result. 
By going back to the original form of the action via 
$\tilde g_{\m\n}=e^{\a\F/2}g_{\m\n}$ 
we get
$$\li {\G_1 =&S-{N-24\over 96\p }\left( \int dx \sqrt {\tilde g}(\tilde R
{1\over {\tilde \Box} }\tilde R - {\a\sp 2\over 4} ({\tilde\nabla} \F )^2 +
\a\tilde R\F )\right) 
+{5\a\over 4\p}
\int dx \sqrt {\tilde g}\tilde R \F \cr 
&-{1\over 2\p }
\int dx \sqrt {\tilde g}
\tilde R{1\over \tilde\Box  }
 (\tilde\nabla \F )^2-{\a \over 4\p }\int dx \sqrt {\tilde g} \F 
 (\tilde\nabla\F )^2 
-{5\a +23\over 8\p }\int dx \sqrt {\tilde g}  (\tilde\nabla \F )^2 \cr
&+ O(e\sp{2\F}) \quad,&(6.11)\cr }$$   
where now S is given by (1.1). In the 
case of large $N$, (6.11) gives 
$$\G_1 =S-{N\over 96\p }\left( \int dx \sqrt {\tilde g}\tilde R
{1\over \tilde\Box }\tilde R-{\a\sp 2\over 4} \int dx \sqrt{\tilde g}
(\tilde\nabla \F )^2+ \a\int dx\sqrt {\tilde g}\tilde R\F \right)
+ O(1/N). \eqno(6.12)$$
In the case of the CGHS model, this is the action proposed
in \cite{bpp}. This action also coincides with the one-loop action
obtained from of the operator quantization of the CGHS model \cite{m96}.

\sect{7. Conclusions}

The results (6.10-11) imply that the leading-order weak-coupling contribution 
to the effective action is independent of the potential $V$. The model 
dependence can be seen in the form (6.11), where it comes from the dilaton
kinetic
term coefficient $\a$ of the original action (1.1). The large-$N$ one-loop
corrections are similar to the BPP case \cite{bpp}, 
but with the model-dependent coefficients. Only the Polyakov-Liouville term
has a model-independent coefficient, which is not surprising, since its
origin is from the integration of the matter fields, whose Lagrangean is
model independent. Note that we have obtained the coefficient
$N-24$ multiplying the Polyakov-Liouville term, which can be understood as 
$c =N+2-26$ conformal anomaly,
where $N$ is the contribution from the matter fields, $2$ is the
contribution from the dilaton and the conformal factor and $-26$ is the
ghost contribution. 

It is interesting that the weak-coupling approximation does not coincide with 
the large-N approximation, given that both approximations are
semiclassical. This is not that surprising, since the
weak-coupling approximation is a more general one, because besides the matter
loops it also includes the graviton and the dilaton loops. 
In the context
of spherically symmetric general relativity this means that 
the large black hole mass approximation (large $N$) is included in the
large radius approximation.
An intriguing feature is that the operator quantization 
of the CGHS model gives the large-N result (6.12)
as the exact one-loop result \cite{m96}, while the covariant perturbation
theory seems to give a non-zero contribution for the lower orders in $N$. 
It is possible that
a resummation of the lower-order in $N$ terms exist, such that it gives zero.
The second possibility is that these two methods give results which 
are not equivalent when $N$ is not large. It would be interesting to 
explore this issue for the case of the spherically symmetric general relativity
as well as for the case of the CGHS two-loop approximation \cite{mr}.

\sect{APPENDIX A}

In this appendix we will compute diagrams which we need for evaluating
$\G_1$. We will use the dimensional  regularization  with $D=2+\e$.
In our case the relevant  lagrangean can be written in the form 
$$\cl=\f^*(\Box +{\overleftarrow \pa}_{a}A^{ab}(x){\overrightarrow \pa}_b  + B_1^b(x)
{\overrightarrow \pa}_b +
{\overleftarrow \pa}_b B_2^b(x) +C(x))\f  \quad,\eqno(A.1)$$
where now $\f$ denotes the set of relevant fields.

The $A$ diagram is given by
$$A =\lim _{x^{\prime}\rightarrow x} 
\int  dxA^{ab}(x) \pa _a \pa ^{\prime}_bG(x-x^{\prime})=
-\int  dx  A^{ab}(x)\int {d^2k  \over (2\p )^2}{k_a k_b\over k^2}.\eqno(A.2)$$ 
(A.2) is infrared divergent  and we must regularize it. We use a modified
minimal subtraction scheme \cite{wgn}, where a counterterm 
is added to the propagator 
$${1\over k^2}\rightarrow  {1\over k^2}-{2i\p ^{{D\over  2}}\over  \e }\d (k)
\G(1- \e/2)\quad. \eqno(A.3)$$
By inserting (A.3) into (A.2) we get $A=0$. Similarly, we get $B=0$.  
The diagram  $C$ is given by  
$$ \int  dxC(x)G(0)=
-\int  dx  C(x)\int {d^2k  \over (2\p )^2}\left({1\over k^2}
-{2\p^{{D\over  2}}i\over  \e }\d (k)\G(1- \e/2)\right)
\quad.\eqno(A.4)$$ 
The first term in  (A.4) is equal to zero, so that
$$C= - {i\p ^{{D\over 2}}\over (2\p )^2 }\G (- \e/2)\int dx C(x).
\eqno(A.5)$$
Therefore $C$ is a UV divergent constant.  

The $AA $  diagram is given by  
$$AA=\int dxdy A^{ab}(x)A^{cd}(y)\pa_{a}^{x} \pa_{d}^{y} G(x-y)\pa_b^x
\pa_c^y G(y-x)
\quad,\eqno(A.6)$$
which is UV divergent. After a suitable Fourier transformation we arrive at
$$\li{AA=&{1\over 4}\int {dp_1\over (2\p  )^4}A^{ab}(p_1)A^{cd}(-p_1)\cr
&\cdot\int dk
{[k_a(p_1+k)_b+k_b(p_1+k)_a][k_d(p_1+k)_c+k_c(p_1+k)_d]\over k^2(k+p_1)^2}
\quad. &(A.7)\cr}$$
This gives 
$$\li {&AA={i\p^{{D\over 2}}\over (2\p )^2}\G(1-\e/2)
B(2+\e/2,2+\e/2){\Big[}-\int dxdy\pa_a\pa_b A^{ab}(x)G(x-y) \pa_c \pa_d
A^{cd}(y)\cr 
+&\int dx \left( -{1\over 2\e (1+{\e  \over 2})}
[A^a_a\Box A^b_b +2 A^{ab}\Box A_{ab}] +{1\over  \e }[ 
3\pa _b A^{ab}  \pa  _dA^d_a-2A^a_a  \pa _c\pa _bA^{bc}]\right) {\Big]} \,.
&(A.8)\cr}$$

The $AB_1$ diagram is given by
$$\li{AB_1 =&\int dxdy A^{ab}(x) B_1^c(y)\pa_b^x G(x-y)\pa_c^y\pa_a^x G(y-x)\cr
=&{i\p^{{D\over  2}}\over  (2\p )^2}B (2+ \e/2,1+ \e/2)
{\Big[} \G(1-\e/2)\int  dxdy \pa _a\pa  _bA^{ab}(x)G(x-y)\pa_cB_1^c(y)\cr 
&-{1\over 2} \G (-\e/2)\int dx\ A^a_a\pa_c B_1^c {\Big]}\,.
&(A.9)\cr}$$
The  $AB_2$ diagram is the same as the $AB_1$ diagram, due to the symmetry
of $A^{ab}$.

The  $AC$ diagram is given by
$$\li{AC = &\int dxdy A^{ab}(x) C(y)\pa_b^x G(x-y)\pa_a^x G(y-x)  \cr
=&{i\p ^{{D\over 2}}\over(2\p )^2}B(1+\e/2,1+\e/2)
  {\Big[}-\G (1- \e/2)\int dxdy \pa_a \pa_b A^{ab}(x)G(x-y)C(y)\cr 
  &+ {1\over 2}\G (-\e/2)\int dx\ A^a_a C {\Big]}\,.
&(A.10)\cr}$$

The $BB$ diagram is given by  
$$\li{BB=\int dxdy {\Big[}&(B_1^a(x)B_1^b(y)+
B_2^a(x)B^b_2(y))\pa_a^x G(x-y)\pa_b^y G(y-x)\cr
+&(B_2^a(x)B_1^b(y)+B_1^a(x)B^b_2(y))\pa_b^y \pa_a^x G(x-y)G(y-x){\Big]}
\,. &(A.11)\cr }$$
If we make a Fourier transformation as before, we obtain 
$$\li{BB=&{i\p^{{D\over 2}}\over (2\p )^2}{\Big[}-\G(1-\e/2)
 B(1+\e/2,1+\e/2)\int dxdy (\pa_a B_1^a (x)\pa_b B_1^b(y)\cr
+& \pa_a B_2^a (x)\pa_b B_2^b (y) -2 \pa_a B_1^a (x)\pa_b B_2^b (y))
G(x-y)\cr
 +&{1\over 2}
\G(-\e/2) B(1+\e/ 2,1+\e/2)\h_{ab} \int dx (2B_2^a B_1^b 
 -B_1^a B_1^b - B_2^a B_2^b)\cr
 -&2\G (1-\e/2)B(\e/2,1+\e/2) 
\int dxdy ( \pa_a B_1^a (x)\pa_b B_2^b (y))G(x-y){\Big]}\,. &(A.12)\cr }$$
From (A.12) we see that the $BB$ diagram is both IR and UV divergent. As
before, we remove the IR divergence by 
the regularization (A.3). By using  
$B(\e/2,1+\e/2)=2/\e+ O(\e )$ 
we obtain
$$\li{BB=&{i\p^{{D\over 2}}\over (2\p )^2}{\Big[}-\G(1-\e/2)
 B(1+\e/2,1+\e/2)\int dxdy (\pa_a B_1^a (x)\pa_b B_1^b (y)  \cr
 +&\pa_a B_2^a (x)\pa_b B_2^b (y) -2 \pa_a B_1^a (x)\pa_b B_2^b (y))G(x-y)\cr
 +&{1\over 2}
 \G(-\e/2) B(1+\e/2,1+\e/2)\h_{ab}\int dx
 (2B_2^a B_1^b -B_1^a B_1^b - B_2^a B_2^b){\Big]}. &(A.13)\cr }$$
The diagrams $BC$ and $CC$ are only IR divergent. After using
(A.3) we get that they are equal to zero. 
In  our  calculations we have
ignored the term $ln{p^2\over \m^2}$, where $\m$ has the dimension of mass.

\end{document}